\documentstyle[12pt,epsf,color,bm]{article}
\newcommand{\be}{\begin{equation}}
\newcommand{\ee}{\end{equation}}
\newcommand{\bi}{\begin{itemize}}
\newcommand{\ei}{\end{itemize}}

\newcommand{\bea}{\begin{eqnarray}}
\newcommand{\eea}{\end{eqnarray}}
\newcommand{\no}{\noindent}

\newcommand{\half}{\frac{1}{2}}
\newcommand{\sign}{{\rm sign\,}}

\newcommand{{\tr}}{\rm tr}

\title{
  {\vspace{-3cm} \Large
      Learning with incomplete information \\
      - and the mathematical 
structure behind it -
  }}

\author{
Reimer K\"uhn$^1$ and
Ion-Olimpiu Stamatescu$^{2}$\\
{\small $^1$King's College, London, United Kingdom}\\
{\small $^2$FESt,  Heidelberg and
 Universit\"at Heidelberg,  Germany}
}

\date{}

\begin{document}

\maketitle

{\bf Learning and the ability to learn are important factors in 
development and evolutionary processes \cite{menz}. Depending on the level,
the complexity of learning can strongly vary. While  associative learning
can explain simple learning behaviour \cite{menz, byrne} 
much more
sophisticated strategies seem to be involved in complex learning tasks.
This is particularly evident in machine learning theory\cite{mit} 
(reinforcement learning
\cite{isut}, statistical learning \cite{vap}),
but it equally shows up in trying to model natural learning behaviour
\cite{byrne}. A general setting
 for modelling learning processes in which statistical aspects are 
relevant is provided by the neural network (NN) paradigm. This is 
in particular of interest for
 natural, learning by experience situations. 
NN learning
models can incorporate  elementary learning mechanisms based on
neuro-physiological analogies, such as the
Hebb rule, and lead to quantitative results concerning the 
dynamics of the learning process \cite{herz}. The Hebb rule, 
however, cannot  be directly applied in all cases, and in particular for
realistic problems, such as ``delayed reinforcement" \cite{isut, herz},
 the sophistication of the algorithms rapidly increases. 
We want to present here a model which can cope with such non trivial
tasks, while still being   elementary and based only on 
procedures which one may think of as  natural, without any  appeal to 
higher strategies \cite{renu}. We can show the capability of this model to 
provide good learning in many, very different settings \cite{renu, mirenu, 
urenu}. It may help therefore understanding some basic features of 
learning. 
}
 
Any realistic form of learning is in some sense learning from experience, 
since a learner interacts with an ``environment", appraises this
interaction and consequently changes its ``internal structure" according to
some criteria. In models of biological 
behaviour, as well as in the design of information processing 
systems, the control for the appraisal procedure -- food, pleasure,
success, assessment of result -- is formalized as some kind of 
{\it reinforcement}. 
Normal ``experience" can, however, rarely be
encoded into a  reinforcement matrix relating actions
with their results in a one-to-one manner, and a learner faces 
the additional task to {\it interpret\/} the environmental feedback before 
rewriting it as an update of its internal (cognitive) structure. 
An urgent problem, for 
example,  with which an ``agent", either natural or artificial, may be 
confronted  is to learn solely from the {\it final}
 success/failure of a series 
of consecutive actions, 
without direct information 
about the particular fitness of each of them, such as learning to 
coordinate
its muscles in moving, 
learning a labyrinth or playing chess.
 Another case  may be that of a feedback which only differentiate between
 the presence or absence of a certain kind of
 patterns in a mixture, such as sorting out unpalatable components
 in food, or realizing
that some actions out of many were useful, without knowing which.
Can sale statistics, for instance, reveal what exactly makes a 
good commercial? 

How non-trivial such problems are can be seen from the sophisticated 
algorithms developed for  so called ``delayed reinforcement"
in the framework of machine learning theories 
\cite{isut, wya}.
In this perspective realistic, complex  learning situations 
seem to require the 
availability of strategies and involved procedures {\it before} 
reinforcement can 
provide effective feedback mechanisms. In an evolutionary perspective of
the development of learning capabilities, 
however, the question then arises as to
how  such strategies and procedures could have developed in the
first place. 
If reinforcement were to be a general or even fundamental element 
involved in effecting behavioural change, one would expect reinforcement 
learning to act {\em in particular\/} at an intermediary level which is 
simple enough not to depend on involved strategies, yet sophisticated enough 
to allow complex behaviour to evolve.

Aside from such basic questions, it should be interesting in itself 
to develop some understanding of the capabilities of {\it simple 
reinforcement learning procedures\/} which do {\em not\/} depend on  
involved strategies to deal with cases of non-specific information.

A  problem of {\it non-specific}
reinforcement can be defined as follows: the reward is global, 
regards the cumulative
result of a series of actions
 and the reinforcement acts non-specifically concerning these actions
\cite{MS85}.
 We ask  whether there may exist
{\em elementary mechanisms} to solve such problems, which may have
developed also under natural conditions and which may hint at basic
features of learning. For this we must not only demonstrate the existence 
of such mechanisms, but also uncover their structural features.

In previous papers 
\cite{renu}, \cite{mirenu}, \cite{lan}
we introduced and studied a learning algorithm for neural networks (NN)
which deals with this problem. The NN setting allows a systematic study 
by both numerical and analytic methods. It provides a framework to study 
learning with non-specific reinforcement
which is transparent, and pertains to the basic machinery on which 
learning is believed to take place. 
NN thus can achieve complex information processing capabilities 
using mechanisms 
simple enough to have plausibly developed under natural conditions.
Due to their general computational capabilities NN can 
model various levels of learning processes -- biological, behavioural or
cognitive \cite{herz}.  
  
The first step of our analysis begins by casting the problem into a 
classification task for a perceptron. In its 
simplest version, this is a network consisting of an  array of  $N$ input 
neurons projecting synapses onto a single output neuron. The active
and inactive states of the neurons are encoded as $+1$ and $-1$ respectively. 
The ``cognitive structure" of the network is encoded in the values $J_i$,  
$i=1,\dots,N$, of the synapse strengths, also called weights. The inputs 
to the network (the ``patterns" which must be classified) are strings of 
$N$ binary values $\xi_i\in \{\pm 1\}$ loaded on the input layer. 
These values are  weighted by the corresponding synapse strengths and 
transmitted to the output neuron, where they are added to define a ``potential"
which, according to its sign, triggers the output neuron to $s=\pm 1$
accordingly attributing the pattern ${\bm \xi}=(\xi_i)$ to the class $s$.

The standard learning problem is stated by asking a ``student" perceptron 
to implement a given classification rule. The rule is provided by a ``teacher"
perceptron with the same architecture, whose synapses $B_i$ are given
and fixed. Student and teacher have access to the same inputs ${\bm \xi}$.
We thus have
\be
s = \sign\left(\frac{1}{\sqrt N} \sum_{i=1}^N J_i \xi_i\right)\ ,\qquad 
 t = \sign\left(\frac{1}{\sqrt N} \sum_{i=1}^N B_i \xi_i\right)\ .
\ee
The student has access to the teacher's output $t$ 
which provides the rule-based classification of the inputs, but {\em not\/} 
to the teacher's rule represented by its weight vector ${\bm B}=(B_i)$. 
The student learns from a stream of inputs classified by the teacher, ${\bm \xi^\mu} 
\to t^\mu$, $\mu=1,2,\dots$, adapting its own weights in response to these data 
in such a way as to reduce the number of cases where its own classification 
${\bm \xi^\mu} \to s^\mu$ disagrees with that provided by the teacher.

A classical, and neuro-biologically motivated learning rule 
is the so-called Hebb rule, 
where synapses change in response to coincidence of 
pre- and post-synaptic activity. 
In the present setting
the appropriate formulation of Hebb's rule for the student performing 
the classification
${\bm \xi^\mu}\to s^\mu$ would take the form
$
\Delta J_i^\mu = J_i^{\mu+1} - J_i^\mu \propto s^\mu \xi_i^\mu .
$

In order to turn this simple form of associative adaptation into {\em 
learning}, it must be supplemented with some form of feedback concerning the 
`quality' of the associations. In learning rules traditionally used in studies 
of supervised learning within the student-teacher scenario the proportionality 
`constant' in Hebb's rule is made to {\em depend\/} on the difference between 
the teacher and student answers. The two most popular algorithms are the supervised 
Hebb (H) and Perceptron (P) learning algorithms, defined by
\bea 
\Delta J_i^\mu =a t^\mu \xi_i^\mu \qquad \mbox{(H)} \qquad  \mbox{and} \qquad 
\Delta J_i^\mu = \frac{a}{2} |s^\mu-t^\mu| t^\mu \xi_i^\mu \qquad \mbox{(P)}\ , \label{e.sh}
\eea
respectively. Written in this form both algorithms assume 
{\em immediate and direct control\/} over 
the student's synapses, which in a way amounts to assuming that a teacher is able 
to clamp a desired output onto the student neuron (e.g., by providing an `evaluating'
stimulus through
some other channel -- see, e.g. \cite{menz}).

In standard ``supervised on-line learning"  the student 
is told after each instance what  the right answer would have been
({\it specific feedback}).
In our approach, however, according to our paradigm the student can 
only receive a {\it non-specific reinforcement} about
some global degree of correctness of its answers {\it over many 
instances}. More precisely
the student is presented with series (``bags") of 
patterns and only obtains information concerning its 
cumulative performance
for the bag as a whole, not with respect to each pattern in the bag.
The non-trivial learning problem is to implement this global
information into a local updating rule for the synapses.
 
The learning algorithm we propose can be described as consisting of two phases. 
To be specific, we assume that each bag $q$ contains the same number $L$ 
of patterns.

In the first phase the student processes the patterns in a bag $q$
one by one and modifies its synapses by simple, 
associative Hebbian learning, 
using its {\em own classifications \/} made
 on the basis of its momentary synapse values:
\bea
{\rm I.\ :}\ \ J_i^{(q,l+1)} = J_i^{(q,l)} + 
\frac{a_1}{\sqrt N} s^{(q,l)}\xi_i^{(q,l)}
\ , \qquad l=1,\dots,L ,
\label{e.a}
\eea
with $s^{(q,l)} = 
 {\sign}\left(\frac{1}{\sqrt N}\sum_{i=1}^N J_i^{(q,l)} 
\xi_i^{(q,l)} \right)$
 being the student classification of pattern $(\xi_i^{(q,l)})$.

 In the second phase the student obtains information about
 its {\it global performance} on a whole bag of patterns
and corrects its synapses by ``reconsidering" the  steps of the 
first phase, and by {\it partially 
undoing them in an indiscriminate way, 
to an extent that depends on the (likewise indiscriminate)
global error over them}, independently on which steps were in 
fact correct and which not. This phase can be seen as Hebbian 
``unlearning":
\bea
{\rm II.\ :}\ \ J_i^{(q+1,1)} = J_i^{(q,L+1)} - 
e_q~\frac{a_2}{\sqrt N} \sum_{l=1}^L 
\omega_l s^{(q,l)}\xi_i^{(q,l)}, \label{e.r}
\eea
where the on-line error $e_q$ is a measure 
of the disagreement to the 
teacher and defines the specific problem (see below). 
In (\ref{e.r}), the  $\omega_l$ can be 1 or 0 with 
probabilities  
$\rho$ and $1-\rho$, respectively,  which accounts for the possibility that the 
replay during
the second phase may be {\em imperfect\/}: the student may not 
recall all associations established during the first phase.

The 
procedure, so to say, is specific but blind association in the 
first phase, 
qualified but  non-specific reinforcement in the second phase.
 We called this algorithm therefore 
{\it Association-Reinforcement (AR) - Hebb} algorithm. 

It is interesting 
to note that a kind of replay as that involved in the 
Phase II of our algorithm apparently can be observed 
in rats on track running tasks
\cite{fowi}. This is considered to correspond to memory 
consolidation. Since experiences are usually not neutral, but
also imply valuations, it is suggestive that in such replay not
only neutral memories are consolidated, 
but memories evaluated by
some measure of success (finding or not 
finding food at the end of 
the track, for instance). This would mean 
observing here a mechanism akin to 
the re-weighting replay of the Phase II of our learning model.
Hebbian unlearning mechanisms via replay of data 
previously exposed to has
been discussed also in other contexts \cite{unlearn}. 
 Concerning 
Phase I this just represent 
strengthening of its own associations by repetition.
Therefore the ``AR-Hebb" 
algorithm not only  appears natural behaviourally, but it may also
have a neuro-physiological basis.

It turns out that only the ratio $\lambda = a_1/a_2$ of learning-rates in phase I 
and II is relevant for the analysis. Hence we shall work mostly with $a_2=1$,
hence $\lambda=a_1$. It measures the strength of the local 
``associative"
compared 
to the global 
``corrective" 
step.  

For a perceptron, the progress of learning depends on the evolution of the 
angle between the student's and the teacher's weight vectors $\bm J$ and $\bm B$. 
At fixed $\bm B$, the relevant quantities are therefore the normalized scalar 
products $R= N^{-1}\bm J\cdot \bm B$ and $Q= N^{-1}\bm J\cdot \bm J$, commonly
referred to as overlaps. Without loss of generality, one can assume the 
normalization $N^{-1}\bm B\cdot \bm B =1$.
For unbiased inputs, the generalization error $\epsilon_G$, i.e., the probability 
of disagreement between student and teacher, can then be expressed in terms of
$R$ and $Q$, $\epsilon_G =\frac{1}{\pi}\arccos(R/\sqrt Q)$. The overlaps $R$ 
and $Q$ are also order parameters in the sense that the dynamics of learning
can at a macroscopic level be fully described in terms of $R$ and $Q$ alone.
The main elements of a derivation of this result are given in {\it Elements of the
analysis} below.

Note that $Q$ controls the relative learning rate (the larger $Q$, the smaller 
the {\it relative} synaptic change induced by a single learning learning step). 

For the case to be presented first, the non-specific reinforcement uses the {\it 
average error} of the student's guesses for the whole bag (this is,  up to 
a normalization, the cumulated, global error). For simplicity of reference we call 
this the ``average error" (AE) problem -- see eq. (\ref{e.AE}). At no moment does 
the student know whether his particular classifications are correct or or incorrect; 
he is only informed about the fraction of correct answers over the whole bag. The 
patterns are produced randomly.

The exciting result of this study is that stable perfect learning can be 
achieved in spite of the non-specific reinforcement! The most
interesting feature is the dependence of learning on
$\lambda$, 
the strength of the associative step as compared with the corrective one. 
It is found that the asymptotic decay of the
generalization error $\epsilon_G$ as a function of the number $q$ 
of pattern-bags used for learning is described by a power-law, $\epsilon_G \sim 
q^{-p}$ with an exponent $p$ that {\em depends\/} on $\lambda$ 
--- an unusual feature in neural network learning. But even more compelling 
is the appearance of a threshold $\lambda_c$ below which {\it no learning 
is possible}, whereas for $\lambda > \lambda_c$ perfect learning is always 
achieved. The exponent $p$ is a decreasing function of $\lambda$, so that 
learning becomes more efficient as $\lambda \searrow \lambda_c$. The value
of $\lambda_c$ itself depends on $L$ and on 
the initial value $Q_0$ chosen for the order
parameter $Q$, that is, on the initial learning rate ($1/\sqrt{Q_0}$). 
{\it There is no such non-zero 
threshold for $L=1$}, where $0 \leq \lambda \leq \half$ just interpolates between 
the supervised Hebb rule (H) at $\lambda=0$ and the Perceptron algorithm (P)
at $\lambda =\half$ in (\ref{e.sh}).

\begin{figure}[htb]
\vspace{8cm}
\includegraphics{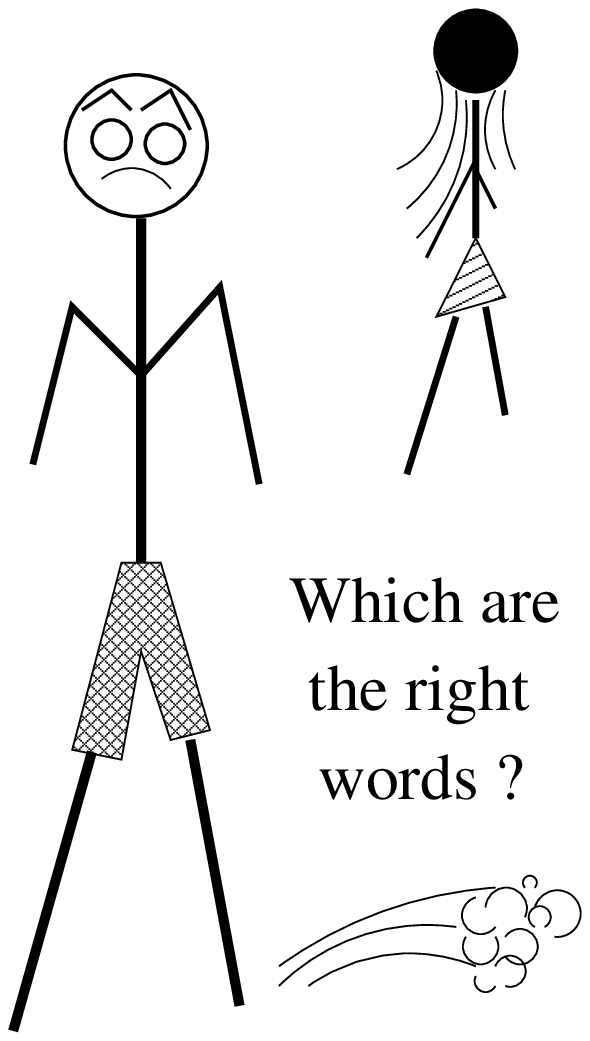}
\includegraphics{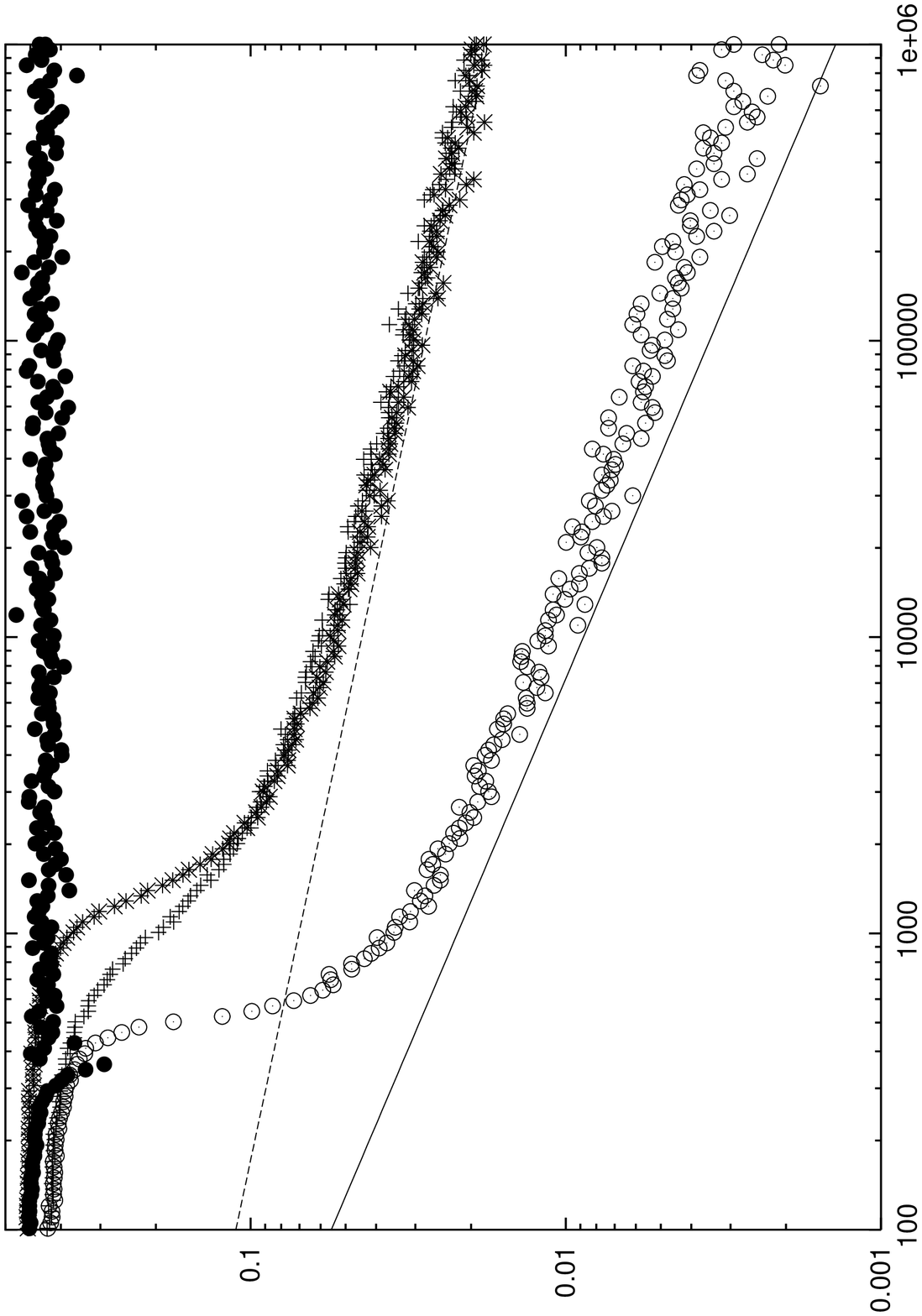}
\caption{Numerical simulation of the AE problem: the 
learner is presented with bags of patterns, feedback is given only via 
the {\it average (or total) error} made over the whole bag. We plot 
the generalization error, 
$\epsilon_G$ vs the normalized number of patterns $\alpha= q L/N$. 
Each of the bags contains $L=10$ patterns; the size of the input layer 
is $N=100$. Initial conditions are random with $R(0)=0$ hence 
$\epsilon_G(0)=0.5$, and $Q(0)=10^4$. The replay phase is complete, i.e.
$\rho=1$ (full circles, empty circles and   
crosses) or only partial ($\rho=0.5$, asterisks).
For $\lambda=0.120$ (full circles)   
no generalization is observed, whereas for  $\lambda=0.125$ (empty circles
and asterisks) and 
$\lambda=0.250$ (crosses) perfect generalization is achievable. 
The lines represent the asymptotic behaviour expected from the analytic 
study $\epsilon_G(\alpha) \propto 
\alpha^{-\frac{1}{2L\lambda}}$. 
}
\label{f.AEsim}
\end{figure}

Fig. \ref{f.AEsim} shows typical results of numerical simulations of the 
learning process with AE feedback for three values of $\lambda$.
For the simulations the generalization error is measured by comparing the 
student and teacher answers on a random set of $10^4$ patterns.

Below $\lambda_c$ (which for the given initial conditions is located 
between $0.120$ and $0.125$) the generalization error initially decreases 
with increasing number $q$ of processed bags, then suddenly returns to a 
value very close to 0.5 and stays there ever after so that generalization 
is very poor in the long run. Just above $\lambda_c$, learning is rapid but 
may be disrupted by finite size fluctuations, which entail that the threshold  
$\lambda_c$ is somewhat fuzzy at finite $N$, becoming sharp only in the 
limit $N\to \infty$, in complete analogy to phase transitions in condensed 
matter systems. Further above $\lambda_c$  finite-size fluctuations cease 
to be effective in disrupting learning, but convergence to perfect 
generalization is also  slower. The plot gives $\epsilon_G$ vs the 
normalized number of processed patterns $\alpha=qL/N$. The straight 
lines indicate the asymptotic behaviour expected for the corresponding 
$\lambda$ from the analytic theory.

Indeed, the results observed in the simulations
can be {\it derived analytically\/} in the limit of large number
$N$ of input-neurons (``thermodynamic limit"), in which the dynamics of 
the order parameters $R$ and $Q$ is shown to be governed by an autonomous 
set of coupled non-linear flow equations. See {\it Elements of the
analysis} below.

This allows  us to proceed to the second step, that of answering structural 
questions. The mathematical structure uncovered in this way  shows
that the global properties of the macroscopic learning dynamics are governed 
by a pair of fixed points, one fully stable, and the other partially stable
with an attractive and a repulsive direction.

\begin{figure}[htb]
\vspace{6.5cm}
\includegraphics{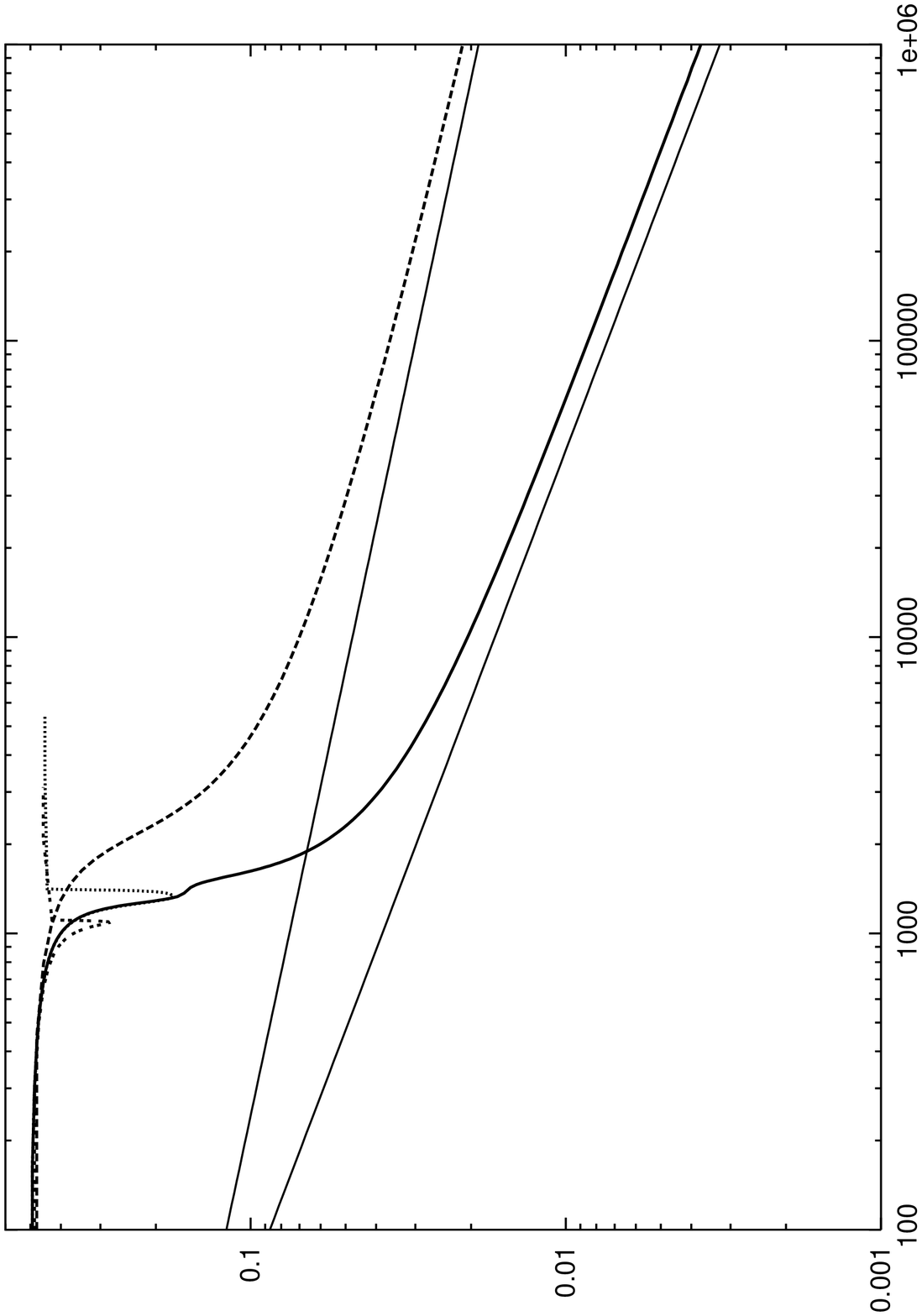}
\includegraphics{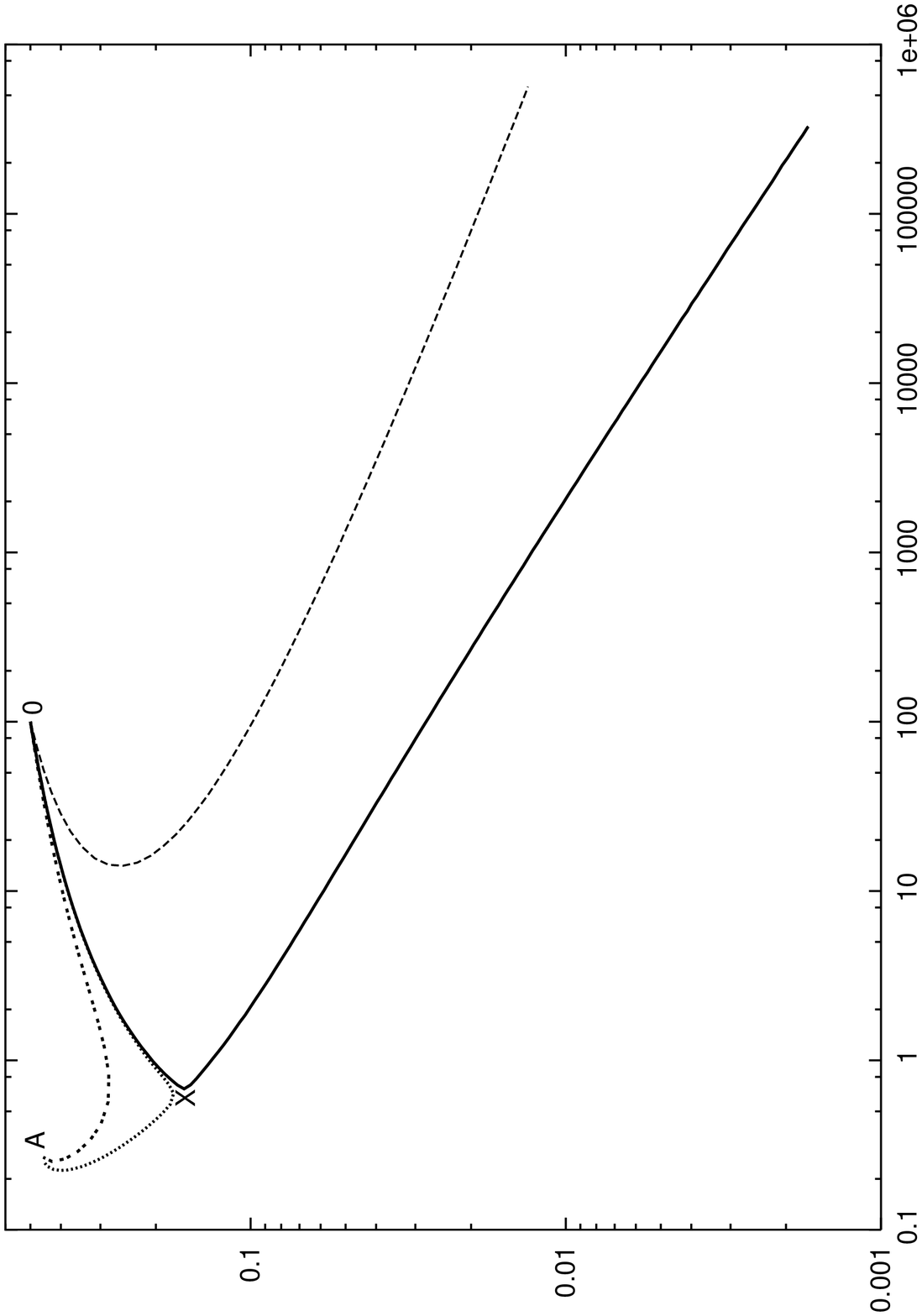}
\caption{Analytic study of the AE problem, $L=10$. Shown are the generalization error
$\epsilon_G$ vs $\alpha$ ({\it left}), and the {\em flow\/} of the learning 
dynamics in the $\epsilon_G,\sqrt{Q}~-$plane ({\it right}) for fixed initial 
condition $R(0) = 0$, $Q(0) = 10^4$, and $\lambda=0.100$ (doubly dotted line),
0.138 (dotted), 0.139 (solid) and 0.250 (dashed line). The threshold 
appears at $\lambda_c \simeq 0.1385$, for which the initial condition is exactly
on the attractive manifold of a partially stable fixed point. 
The attractive fixed point (A) and the
partially stable fixed point (X) are clearly identifiable. The initial condition
is marked with a $0$ in the plot. The straight lines on the left plot represent 
the asymptotic power laws with $p=\frac{1}{2L\lambda}$.}
\label{f.AECG}
\end{figure}

In Fig. \ref{f.AECG} we present analytical results for the learning process 
as obtained by solving the flow equations (\ref{e.Rdyn}), (\ref{e.Qdyn}). 
The left panel is to be compared with the simulation
results shown in Fig. \ref{f.AEsim}. In the right panel, which shows the 
phase-flow of the learning dynamics in the $\epsilon_G,\sqrt{Q}~-$plane, 
we can clearly discern the existence of a separatrix connecting the starting 
point and a partially stable fixed point with an attractive and a repulsive 
direction. The repulsive manifold directs the flow of the learning dynamics 
either towards large $Q$ and perfect generalization, or towards small $Q$
and an all-attractive fixed point of poor generalization. The alternative 
is decided by $\lambda$ which determines on which side of the separatrix 
the initial condition finds itself. By changing $\lambda$ as we do here, we
actually move the fixed points around, and continuously deform the repulsive and
the attractive manifolds, thereby sweeping the separatrix across the initial 
condition. Exactly at $\lambda_c$ the initial condition is found to lie 
{\em on\/} the  separatrix. The structure indicated here fully explains the 
behaviour described above. The thresholds are crisp, of course, as 
there are no longer any finite size fluctuations. By comparing with 
results shown in  Fig. \ref{f.AEsim} the finite-size correction of 
$\lambda_c$ at  $N=100$ is found to be approximately 15\%.

The mathematical structure lying behind our results suggests that  
they are stable against changing parameters and further details of the 
learning process, and in fact may hint at some general properties of our learning 
paradigm. A number of further studies were performed in order to 
substantiate this suggestion.

First, we investigated what happens if the ``replay" in 
Phase II is not perfect --- as a way to describe the possibility
that the student may not `remember' all  instances  encountered 
during Phase I. We modelled this by randomly including each instance 
of Phase I only with probability $\rho \leq 1$ in the unlearning step 
of Phase II. It turned out that all features observed for
complete replay are preserved; for the asymptotic domain of good 
generalization the modification amounts to a re-scaling 
of the learning parameter  $\lambda \to \lambda/\rho$ \cite{renu}
 -- see Fig. \ref{f.AEsim}.

Another possibility concerns randomly varying bag sizes $L$ used in 
learning. These, too, lead to results qualitatively unchanged when 
compared to the case of fixed $L$.

The next modification concerns the nature of the on-line error $e_q$
used in the reinforcement phase. We investigated a case where
the student is only told the {\it number\/} of patterns of, say, 
class $+1$ in a bag, which he or she can  compare with the number perceived to belong to 
that class by him/herself -- see eq. (\ref{e.HI}). We called this 
the ``hidden instance 
problem" (HI); it is a version of the so called ``multiple instance 
problem". The information received by the student thus has an increased 
degree of non-specificity (since, e.g., $e_q = 0$ does not yet 
imply that the student has found 
the correct classification!). Nevertheless the performance 
of our learning 
model for this problem is very similar to that for the AE-problem.
This can be seen both in simulations and in the corresponding analytic theory. 
The quantitative details are different, but the general behaviour is the 
same: Once more, we find a threshold $\lambda_c$, below which generalization 
remains poor in the long run, and a $\lambda$-dependent asymptotic power-law 
decay of the generalization error above the threshold \cite{urenu}.

A further extension looked into using structured input data in the 
classification task to be learnt, as a highly schematic way to 
model learning in a structured environment. Once more, similar
results were obtained \cite{mirenu}. However, some supplementary 
tuning of the learning rate $\lambda$ was necessary, as indeed for the
corresponding {\em standard\/} supervised algorithms dealing with the 
same problem. The dynamics itself is more complex, requiring {\em three\/}
order parameters for a full macroscopic description instead of two. 
Yet again we observe that the parameter $\lambda$  must exceed a 
threshold $\lambda_c$, which depends on initial conditions, in order
to achieve asymptotically perfect generalization.

\begin{figure}[htb]
\vspace{8cm}
\includegraphics{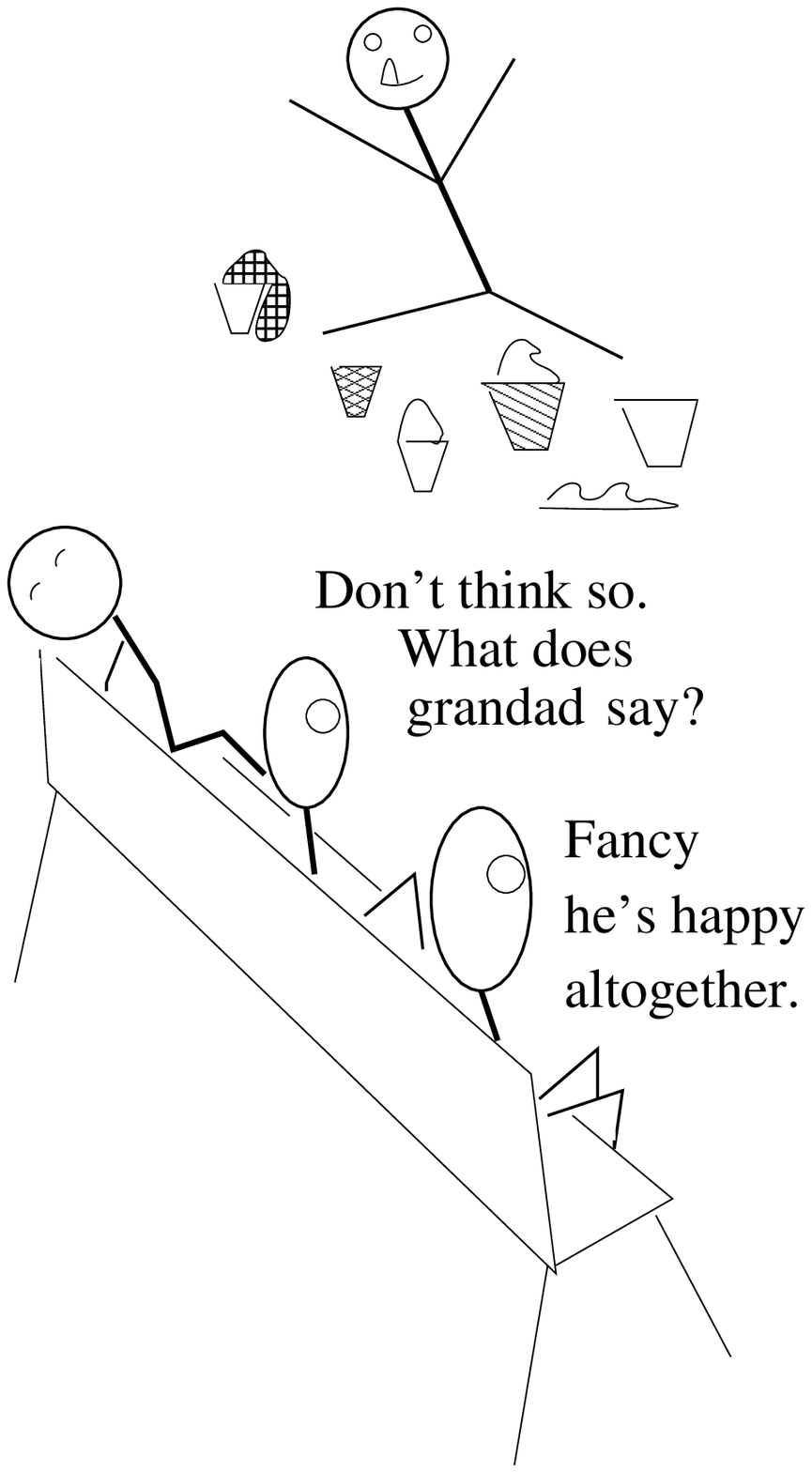}
\includegraphics{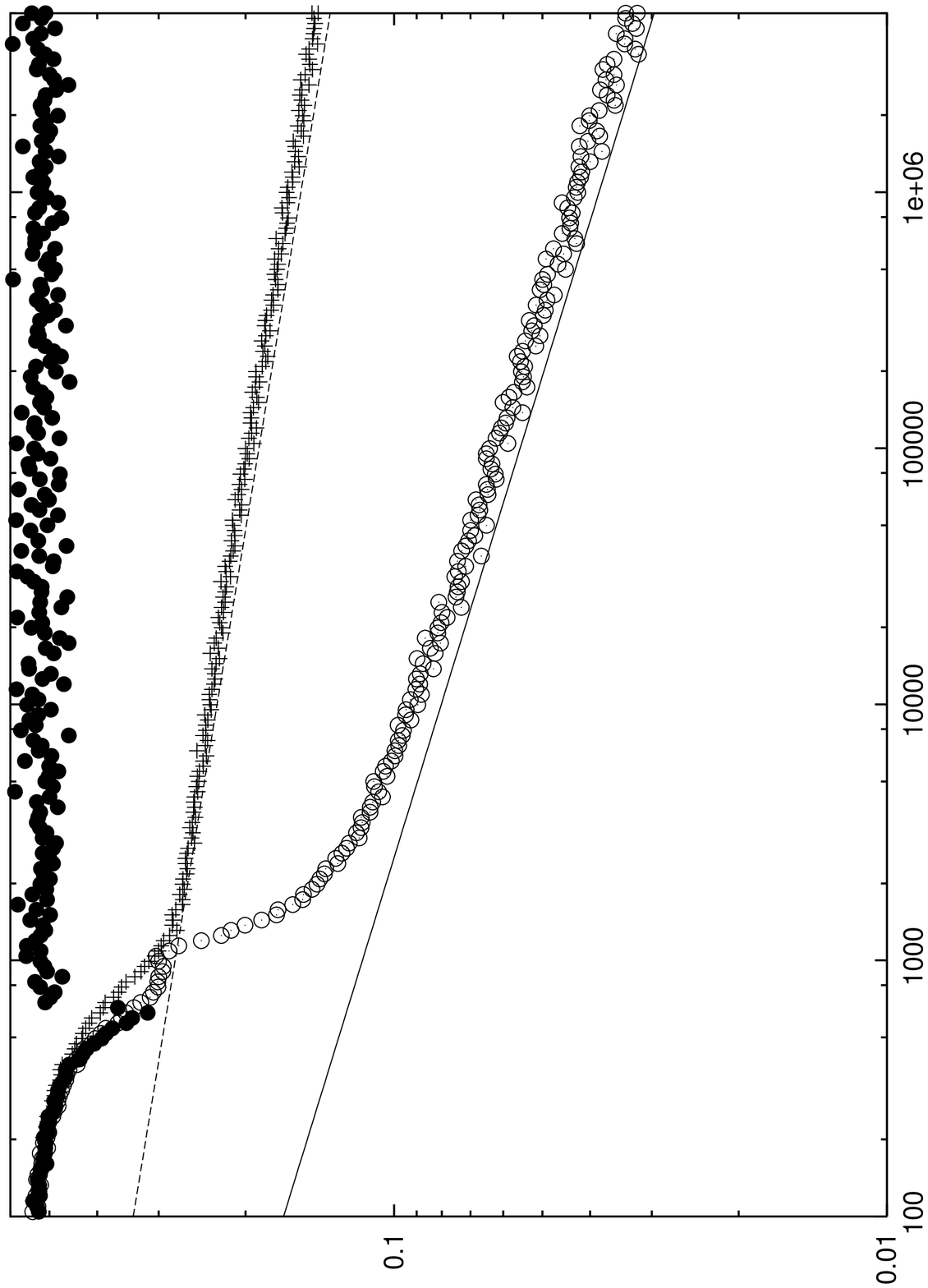}
\caption{Numerical simulation for a committee machine
 with $K=3$ hidden units and $N=100$ input units, AE  problem 
with $L=5$: generalization error $\epsilon_G$  vs $\alpha$.
Here $\lambda=0.260$ (full circles), 0.280 (circles) and 0.320 (crosses).
The straight lines represent power-law fits to the asymptotic
behaviour with exponents $p=0.16$ (solid line) and $p=0.085$ (dashed
line), respectively.}
\label{f.HI-com}
\end{figure}

In the cases discussed above we investigated single-layer networks, 
for which the solvable classification tasks are limited to the class 
of so-called ``linearly separable" problems. This means that problems
in a different complexity class, and presumably some of the more realistic 
ones, can at first sight not be solved by such networks, irrespectively 
of the learning algorithm used to train them, 
while they can be attacked by
multi-layer networks.  

It is now well known that the limits of linear separability can be
transcended by the use of preprocessing and kernel methods \cite{schoe}, 
so as to provide the capability for universal classification while adhering 
to the perceptron as the trainable neural element.

Nevertheless, in order to lend further credibility to the hypothesis that 
non-specific reinforcement could provide a basic learning mechanism at work 
also {\em beyond the single neuron level\/}, and thereby plausibly contribute 
to the evolution of complex information processing capabilities in neural 
architectures, we investigate its performance on a simple multi-layer network
\cite{urenu}, the so called ``committee machine". This is a two-layer network 
with the neurons of the second (hidden) layer -- the committee -- transmitting 
their state via fixed  synapses to the output neuron. Only the synapses from 
the input neurons to  committee members can be modified in the learning process. 

There is an important second motivation, beyond that of demonstrating the
viability of the non-specific reinforcement principle for training simple
multi-layer networks, capable of performing classifications outside the 
linearly separable class: It is related to the fact that the single output of 
a multi-layer networks is {\em itself non-specific\/} in the sense of not 
revealing which of possibly several states of the hidden layer was responsible
for it. Specifically, in the case of a committee-machine producing a simple
majority vote of the committee members, no information is revealed as to 
which subset of the committee was backing the majority vote. This is 
non-specificity with respect to contributions of hidden nodes (for simplicity
referred to as `non-specificity with respect to space'), whereas the {\em AR
- Hebb\/} algorithm introduces an element of non-specificity with respect 
to time. By using the {\em AR - Hebb\/} algorithm to train a committee machine, 
we {\em combine\/} non-specificities in space {\em and\/} time, and the natural
question arises whether this further reduction of the information used for
feedback still permits that a rule --- represented by a teacher committee of
the same architecture --- can be picked up on the basis of classified inputs
alone.

We are able to report here recent results about this system. The version
we have looked at is a ``graded" version producing as its output the sum
of the outputs of all committee members, without performing a final 
sign-operation on that sum. The simulations do indeed
show convergence to perfect generalization, and a threshold in the
learning parameter $\lambda$, as for the perceptron. See Fig.
\ref{f.HI-com}. However, this turns out to be combined with an even 
more complex picture of the evolution of the order parameters, 
hinting at a more complicated fixed point structure in the phase flow. 
E.g., a partially stable fixed point representing the student committee
in a state where its hidden nodes have not yet specialized to represent
one of the hidden nodes of the teacher committee  exists \cite{saso}
which interferes with the fixed point specifically associated with the 
unspecific delayed reinforcement. The interference between fixed points
entails that effects of finite size fluctuations are stronger.   
The analytic study of the algorithm in terms of flow equations involves
more order parameters (the mutual overlaps between weight vectors of all 
combinations of hidden nodes of student and teacher committee) and is 
is thus more difficult, but a number of special results strengthen the 
findings from the numerical simulations. 

To summarize, a simple learning algorithm based on local Hebb-type 
synaptic modifications can solve various non-specific reinforcement  
problems. The  algorithm seems applicable to a broad range of different 
situations, which hints at a high level of generality.  It is interesting 
that such a simple algorithm can cope with complex learning problems,
and this makes it a candidate for a basic mechanism in learning. 
As a model for biological developments it indicates that feedback 
non-specificity can be dealt with at the elementary neuronal level
by mechanisms which are simple enough to have plausibly developed
during the early stages of evolution. A very peculiar aspect is the 
essential role of {\it both}, local Hebb potentiation and global 
correction via replay. 

Note that the correction in phase II of the
 algorithm respects an
important informa\-tion-theoretic symmetry: the synaptic correction
in response to an indiscriminate feedback is indiscriminate {\em
in the same sense}. Any deviation from this symmetry would entail
that the learning mechanism creates a hypothesis about its environment
for which the environmental feedback does not provide any evidence.
This is in fact in fairly close analogy with Bernoulli's principle
of insufficient reason according to which the best assumption in an
information-theoretic sense about a random variable of which we have
no knowledge whatsoever apart from the range of values it can take,
is to  assume that its distribution is {\em uniform\/} over the range
of possible values.

Some features of the learning behaviour described by this model may also 
show up in more complex non-specific feedback situations. In a
behavioural setting, for instance, we find commitment to one's own 
experiences  and global, critical 
consideration of the results as 
necessary prerequisites for learning under the
conditions of non-specific feedback.
In the interaction between
 these two factors
the randomness of experiences is shaped into a knowledge landscape.
Nevertheless in our eyes the first 
merit of this model is to provide an
elementary mechanism for learning from experiences with non-specific 
feedback which may be relevant in an evolutionary perspective.

% \par\bigskip

\newpage
\no {\it Elements of the analysis}

The synapse from the $i$-th input neuron ($i = 1,\dots,N$) onto the output 
neuron 
is denoted $J_i$ for the student, $B_i$ for the teacher. The former 
change under learning, the latter are fixed (randomly given). 
The patterns in the bag $q$ 
are denoted 
$\xi_i^{(q,l)}$, $l = 1,\dots,L$, they are random series of $\pm 1$'s. 
The output of the student and of the 
teacher for a given pattern are, respectively:
\bea
s^{(q,l)} = {\rm sgn}\left(\frac{1}{\sqrt N}\sum_{i=1}^N J_i^{(q,l)} \xi_i^{(q,l)} \right) ,\quad
t^{(q,l)} = {\rm sgn}\left(\frac{1}{\sqrt N}\sum_{i=1}^N B_i\xi_i^{(q,l)}\right) .\label{e.ast}
\eea
\no The teacher synapses are normalised to 
$\frac{1}{N} \sum_{i=1}^{N} B_i^2 = 1$. By rescaling the student synapses 
with $a_2$, $J_i^{(q,l)}/a_2 \to J_i^{(q,l)}$, we remain with  $\lambda = a_1/a_2$ as 
the only relevant learning parameter.

The two phases of the updating algorithm are then
\bea
J_i^{(q,l+1)} &=& J_i^{(q,l)} +\frac{\lambda}{\sqrt N}  s^{(q,l)}\xi_i^{(q,l)}\ , 
\qquad l=1,\dots,L \label{e.ea} \\
J_i^{(q+1,1)} &=& J_i^{(q,L+1)} - e_q \frac{1}{\sqrt N}~\sum_{l=1}^L \omega_l
s^{(q,l)}\xi_i^{(q,l)}\ . \label{e.er}
\eea
For the remainder of the present analysis we shall for simplicity restrict ourselves
to the case where replay during phase II is complete, i.e., $\omega_l\equiv 1$.

We define the ``average error" (AE) problem by the 
following global return (error):
\bea
e_q^{AE}  = \frac{1}{2L} \sum_{l=1}^L \left|s^{(q,l)}-t^{(q,l)} 
\right|. \label{e.AE} 
\eea
which measures the fraction of inputs in current bag, on which student and teacher 
disagree.
For another problem, referred to  in the text as 
``hidden instance"
(HI) problem the global return is
\bea
e_q^{HI}  =  \frac{1}{2L} \left|\sum_{l=1}^L s^{(q,l)}- 
\sum_{l=1}^L t^{(q,l)}\right| . \label{e.HI} 
\eea
measuring the discrepancy between student and teacher 
concerning the  current
balance between positively and negatively classified input data in a bag.

The quantity of interest is 
$\epsilon_G$, the ``generalization error" 
which measures the probability of disagreement
between the teacher and the student on a random set of patterns.
The generalization error can be expressed in terms
of the angle between the weight vectors of student and teacher. 
A simple geometrical
argument \cite{OppKiKlNe} gives 
$\epsilon_G = \frac{1}{\pi} \arccos ({\bm J};{\bm B})$.

In terms of the normalized scalar product of the student's  
and teacher's weight vector and of a correspondingly normalized scalar product 
of the student's weight vector with itself 
\bea
R_q &=& \frac{1}{N} {\bm J^{(q,1)}}\cdot {\bm B}=
\frac{1}{N} \sum_{i=1}^{N} J_i^{(q,1)}\,B_i\ , \label{e.e}\\
Q_q &=& \frac{1}{N}  {\bm J^{(q,1)}}\cdot {\bm J^{(q,1)}}=
\frac{1}{N} \sum_{i=1}^{N} \left(J_i^{(q,1)}\right)^2, \label{e.q}
\eea
respectively, one obtains
\bea
\epsilon_G(q)= \frac{1}{\pi} \arccos ({\bm J^{(q,1)}};{\bm B}) 
= \frac{1}{\pi} \arccos \left( \frac{R_q}{\sqrt{Q_q}}\right)
\eea
at the beginning of session $q$.

For the present system the quantities $R_q$ and $Q_q$ are 
also ``order parameters" in 
the sense that the dynamics of learning can in the large-$N$ 
limit be described {\em in 
terms of these macroscopic quantities alone}.

From (\ref{e.a}),(\ref{e.r}),(\ref{e.e}),(\ref{e.q}) 
one obtains, on combining the 
{\em total effects\/} of phase I
and phase II learning of a bag $q$,
\bea
R_{q+1}&=& R_q  + \frac{\lambda - e_q}{N}
 \sum_{l=1}^L \sign\left(x^{(q,l)}\right) y^{(q,l)}
\label{e.Rupd}\\
Q_{q+1}&=& Q_q  + 2 \frac{\lambda - e_q}{N} 
\sum_{l=1}^L \sign\left(x^{(q,l)}\right) x^{(q,l)}  + 
L\frac{(\lambda - e_q)^2}{N}\ .\label{e.Qupd}
\eea
Here we have introduced
\be
x^{(q,l)} =\frac{1}{\sqrt N}\sum_{i=1}^N J_i^{(q,l)} \xi_i^{(q,l)} \ , \qquad 
y^{(q,l)} =\frac{1}{\sqrt N}\sum_{i=1}^N B_i \xi_i^{(q,l)} \label{e.fields}
\ee
and exploited the fact that $N^{-1} \sum_i \xi_i^{(q,k)} \xi_i^{(q,l)} = \delta_{k,l}$ in 
the large $N$ limit. The remainder of the analysis consists (i) in introducing `continuous 
time' $\alpha = qL/N$, so that $R_q \to R(\alpha)$, and $Q_q \to Q(\alpha)$, (ii) in
realizing that the central limit theorem entails that the fields $x^{(q,l)}$ 
and $y^{(q,l)}$ are zero-mean Gaussian with correlations that depend {\em only\/} 
on $R(\alpha)$ and $Q(\alpha)$,
and (iii) in combining a large number of updates (\ref{e.Rupd}),
(\ref{e.Qupd}) to obtain an autonomous pair of ODEs, which can
be formulated in terms of {\em averages\/} over these updates,
and which describe the learning dynamics in the large $N$ limit
\bea
\frac{{\rm d}R}{{\rm d}\alpha} &=& \left\langle \frac{\lambda - e_q}{L} 
\sum_{l=1}^L \sign\left(x^{(q,l)}\right) y^{(q,l)}\right\rangle
 \label{e.Rdyn}\\
\frac{{\rm d}Q}{{\rm d}\alpha} &=& 2 \left\langle \frac{\lambda - e_q}{L} 
\sum_{l=1}^L \sign\left(x^{(q,l)}\right) x^{(q,l)} \right\rangle + 
\Big\langle (\lambda - e_q)^2 \Big\rangle
 \label{e.Qdyn}
\eea
The angled brackets in (\ref{e.Rdyn})and (\ref{e.Qdyn}) denote averages over the Gaussian variables
$x^{(q,l)}$ and $y^{(q,l)}$ (which are uncorrelated for different indices $l$) and can be 
evaluated in terms of $R(\alpha)$ and $Q(\alpha)$ alone \cite{renu, urenu}.

The numerical simulations implement the network operation (\ref{e.ast}) and
the learning rules (\ref{e.ea}),(\ref{e.er}) at the microscopic level. The analytic 
study is based on the ODE's (\ref{e.Rdyn}) and (\ref{e.Qdyn}) which describe network
performance at a macroscopic level. They can be
solved  numerically at all $\alpha$ and sometimes also analytically for large $\alpha$  
\cite{renu, urenu} (see also \cite{mirenu} for the case of structured data).
They reveal the fixed point structure that directs the flow as discussed in the text.

We close with two {\em illustrations\/} to demonstrate that the learning algorithm 
proposed above and studied in the context of formal neural networks can be applied
to solve ``real world" problems.

 \par\bigskip
\no {\it Illustration 1}

The following simulation is intended to illustrate the above model for a
``realistic" problem: an agent moving on a board with obstacles must learn
that it is good to reach the upper line and how to find its way there. 
The board is partitioned into a regular grid of squares
%%% in equal cells 
and the agent takes
one step at a time (up, down, left or right). 
It receives a positive or negative reward at the end of the journey, proportional to the number of steps it took to reach the upper line
(it always starts at the middle of the bottom line). The ``cognitive structure" 
of the agent is realised as a network with 20 input neurons (storing the 
information {\it free/occupied} concerning the neighbouring cells for the last 
5 steps) and a ``committee" of 4 neurons, each responsible for one of the
4 directions of move. The winner is chosen with probability 
$p_k = {\rm e}^{\beta h_k}/\sum_{k'=1}^4 {\rm e}^{\beta h_{k'}}$, where
 $h_k$ denotes the activation potential of the neuron representing
 direction $k$,
 and $\beta$ is a parameter allowing to vary 
the degree of randomness. 
The synapses (weights) from the input layer to the
committee are modifiable.
Learning proceeds along the lines described above: 
an immediate Hebb reinforcement of the weights after each step and a 
readjustment  at the end of a path using the global information on the
total number of steps (trying to run against obstacles implies an immediate Hebb-penalty). 
The model is described in more detail in \cite{lan}, 
here we only briefly illustrate its performance. 

This problem is of the type AE (average error) above, with a 
journey representing a ``bag" of decisions. It involves however 
a strong mutual dependence of the local decisions (since different moves may
lead to different later situations).
Interesting features of the learning behaviour are:\par
--- good paths are found very fast, without needing any built-in strategies, and
in spite of the fact that feedback is given only via global reinforcement, 
 \par
--- hierarchical learning is possible - new behavioural rules (the direction of move
in a certain situation) are added without discharging old ones (if not
contradictory), \par
--- the randomness employed in choosing the direction of moves helps 
optimising the behaviour, or coping with changes in the situation, \par
--- stable behaviour is compatible with fluctuations, \par
\begin{figure}\protect
\vspace{6cm}
\includegraphics{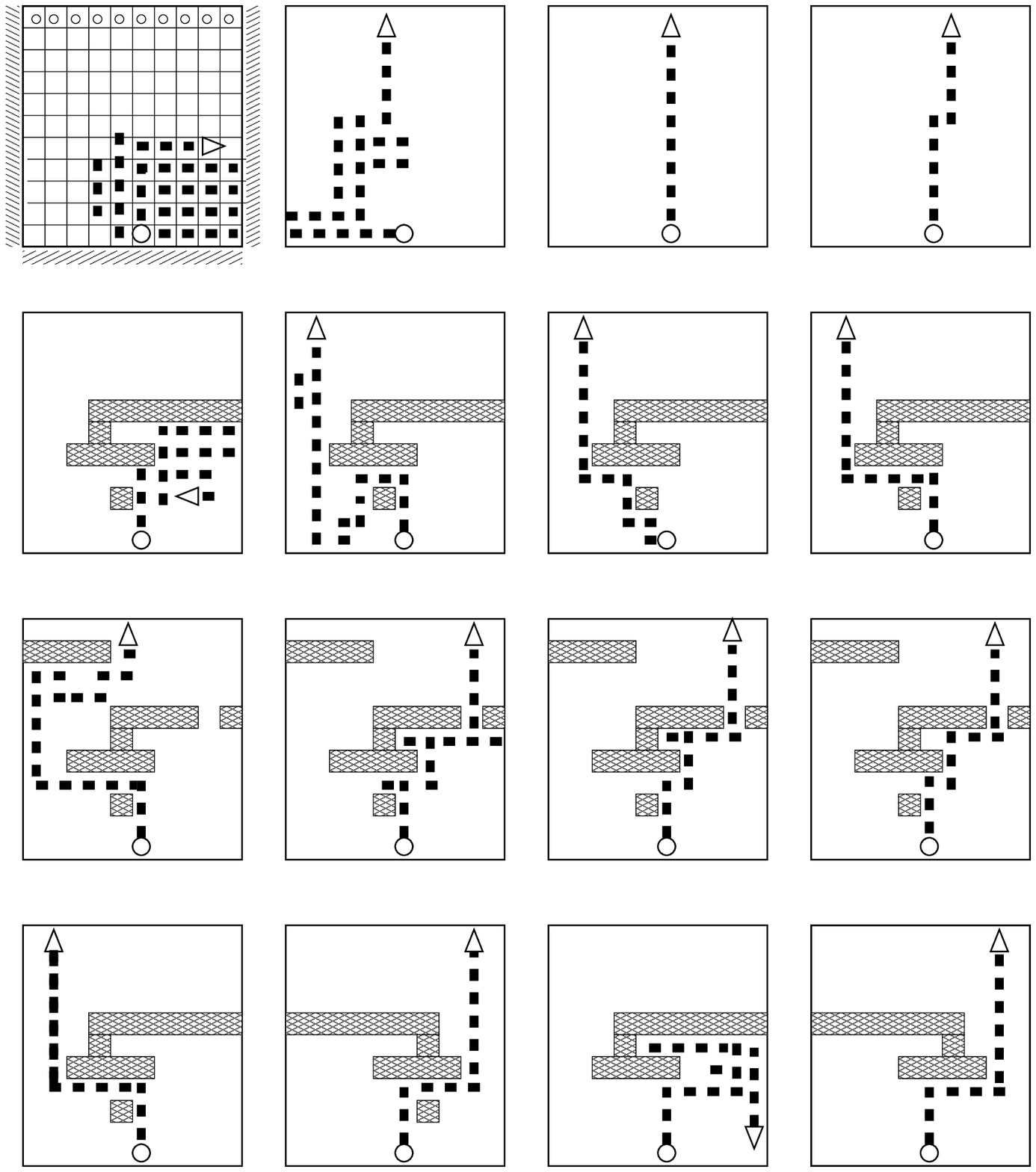}
\includegraphics{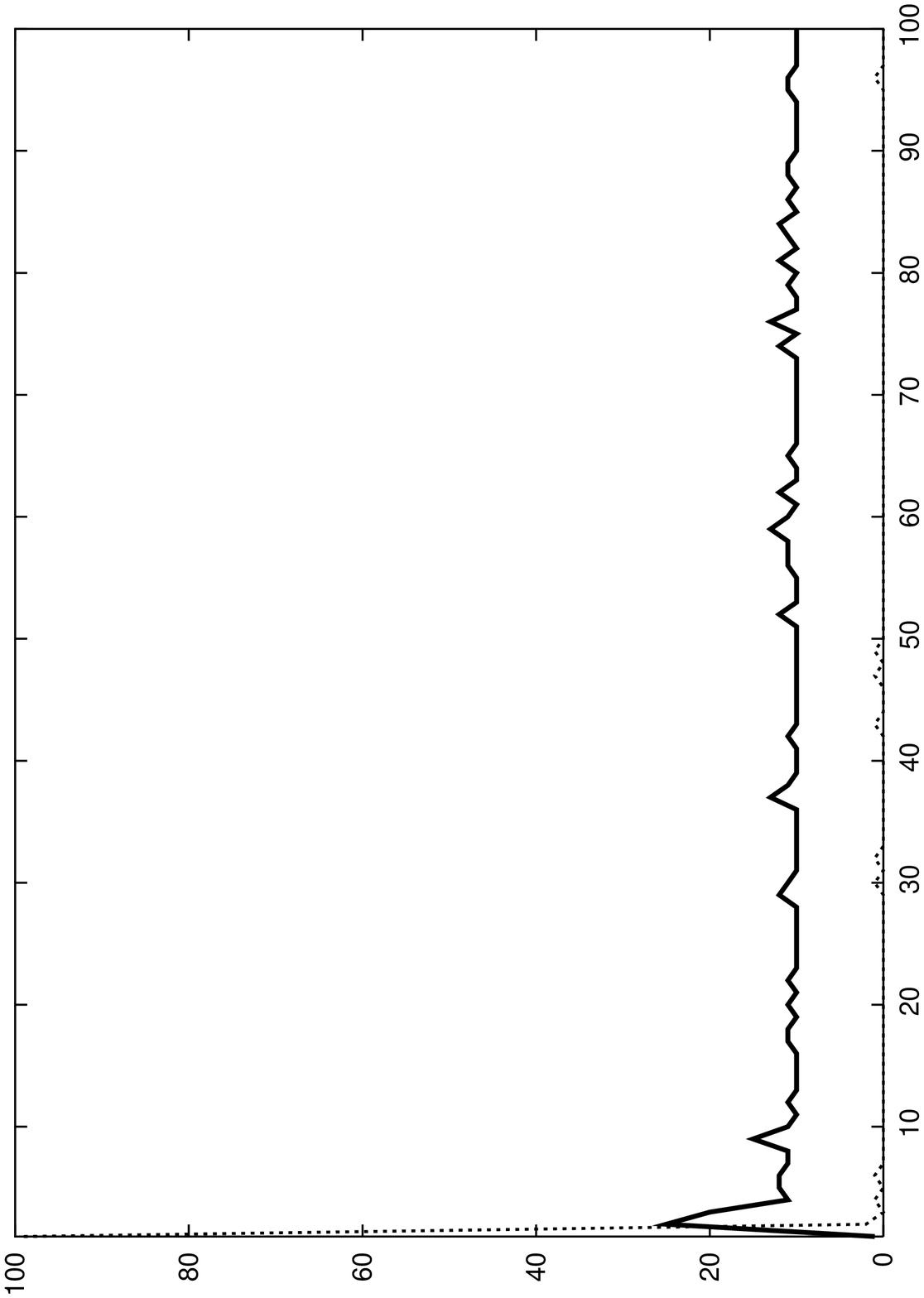}
\includegraphics{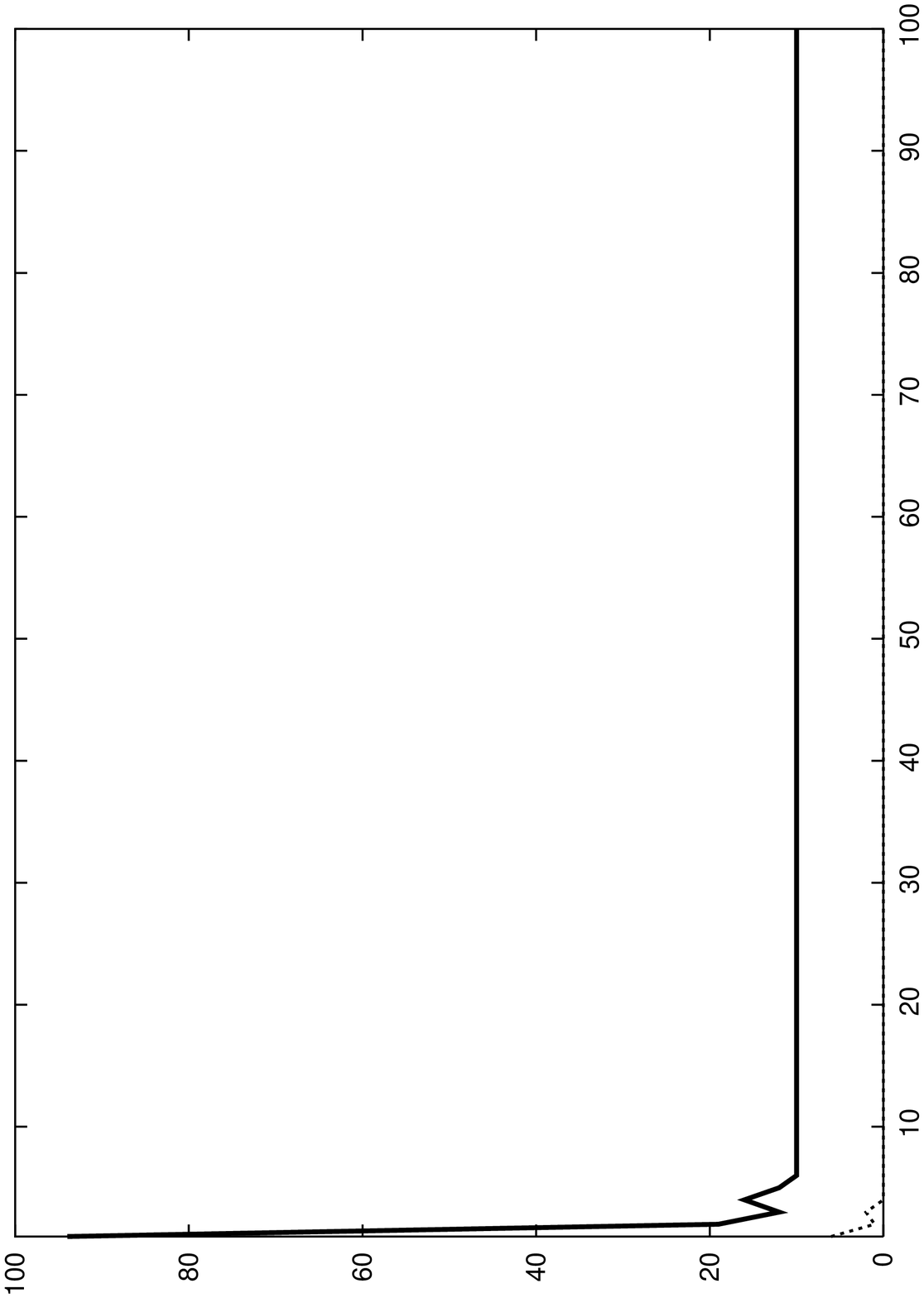}
\includegraphics{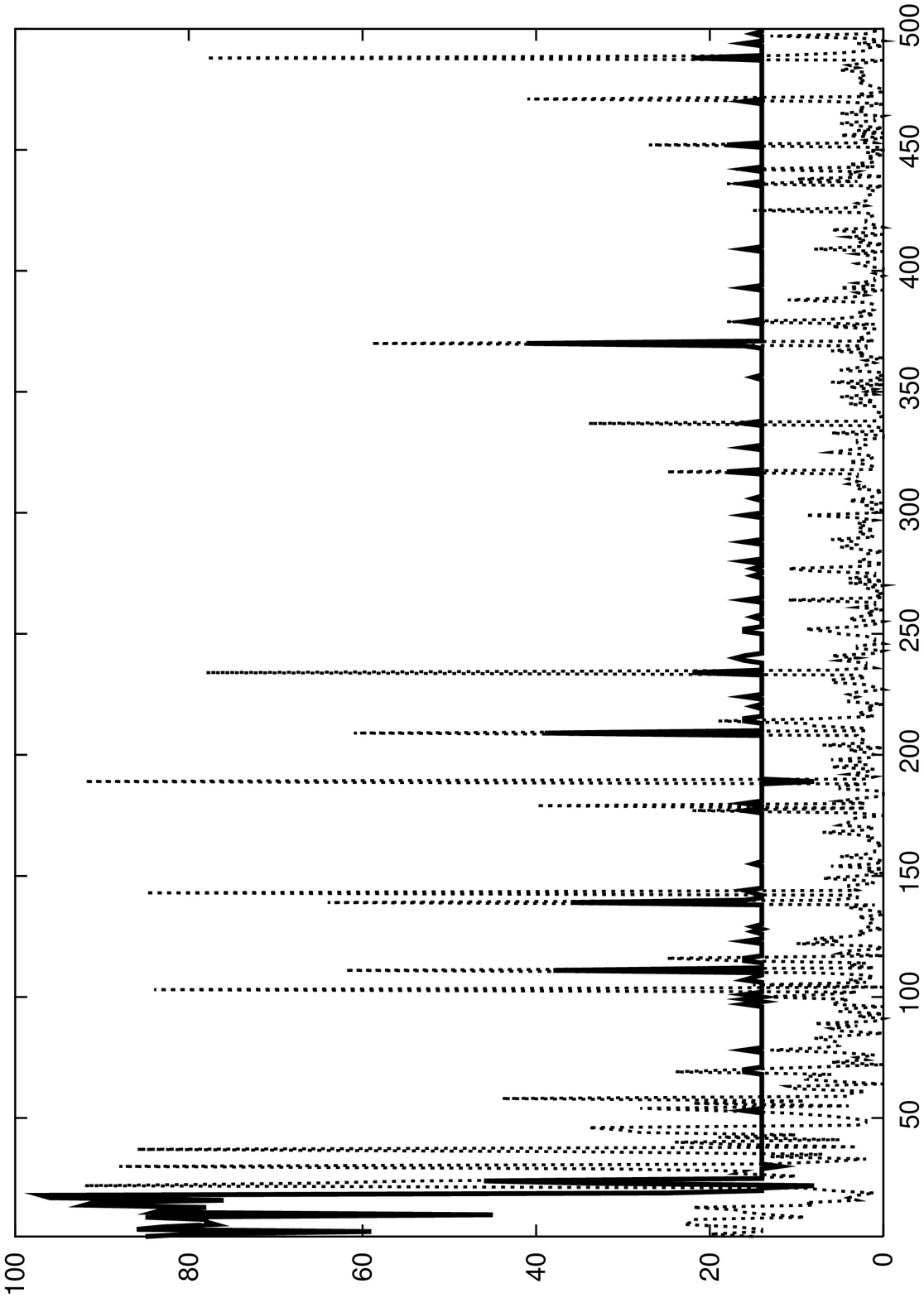}
\includegraphics{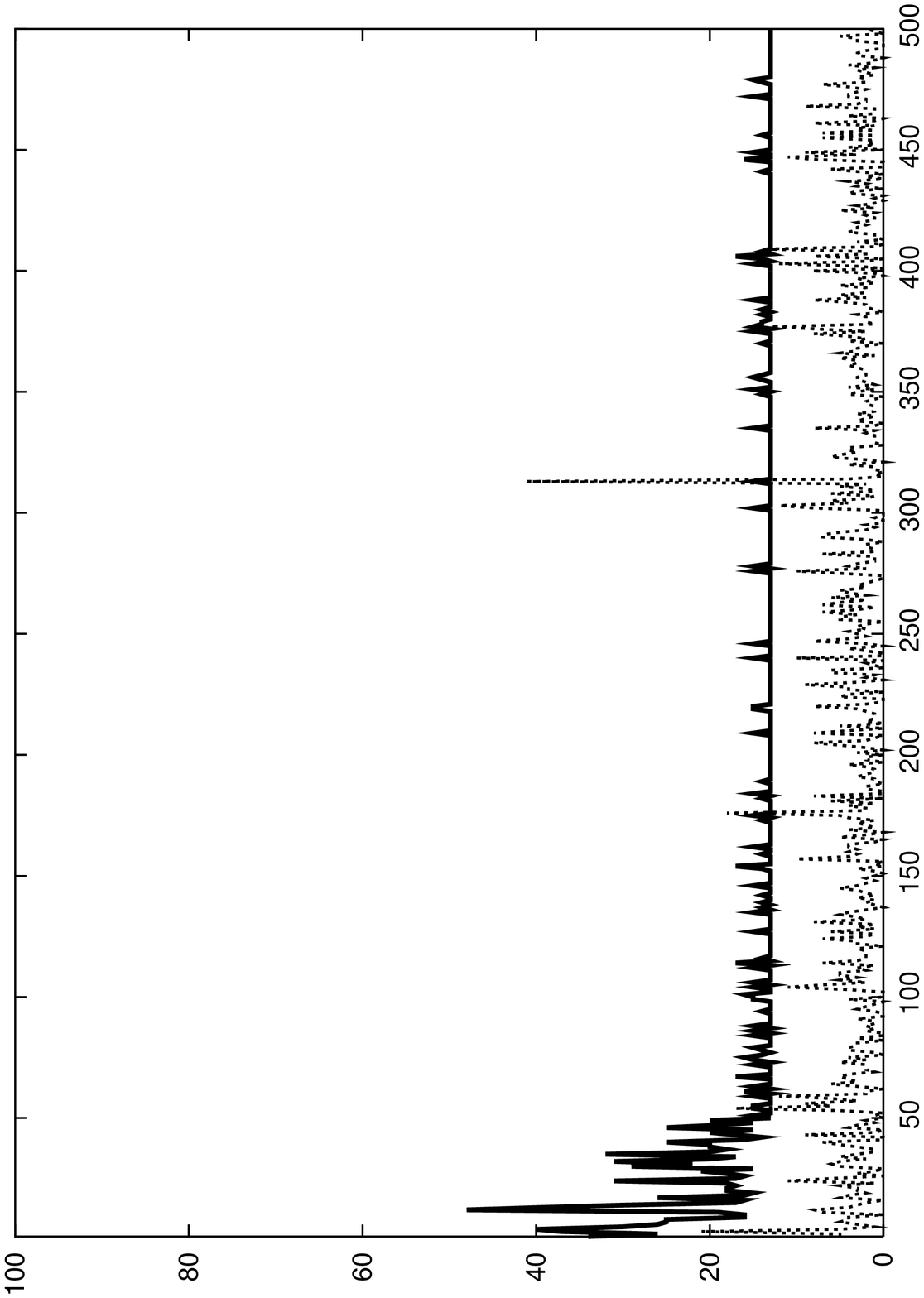}
\includegraphics{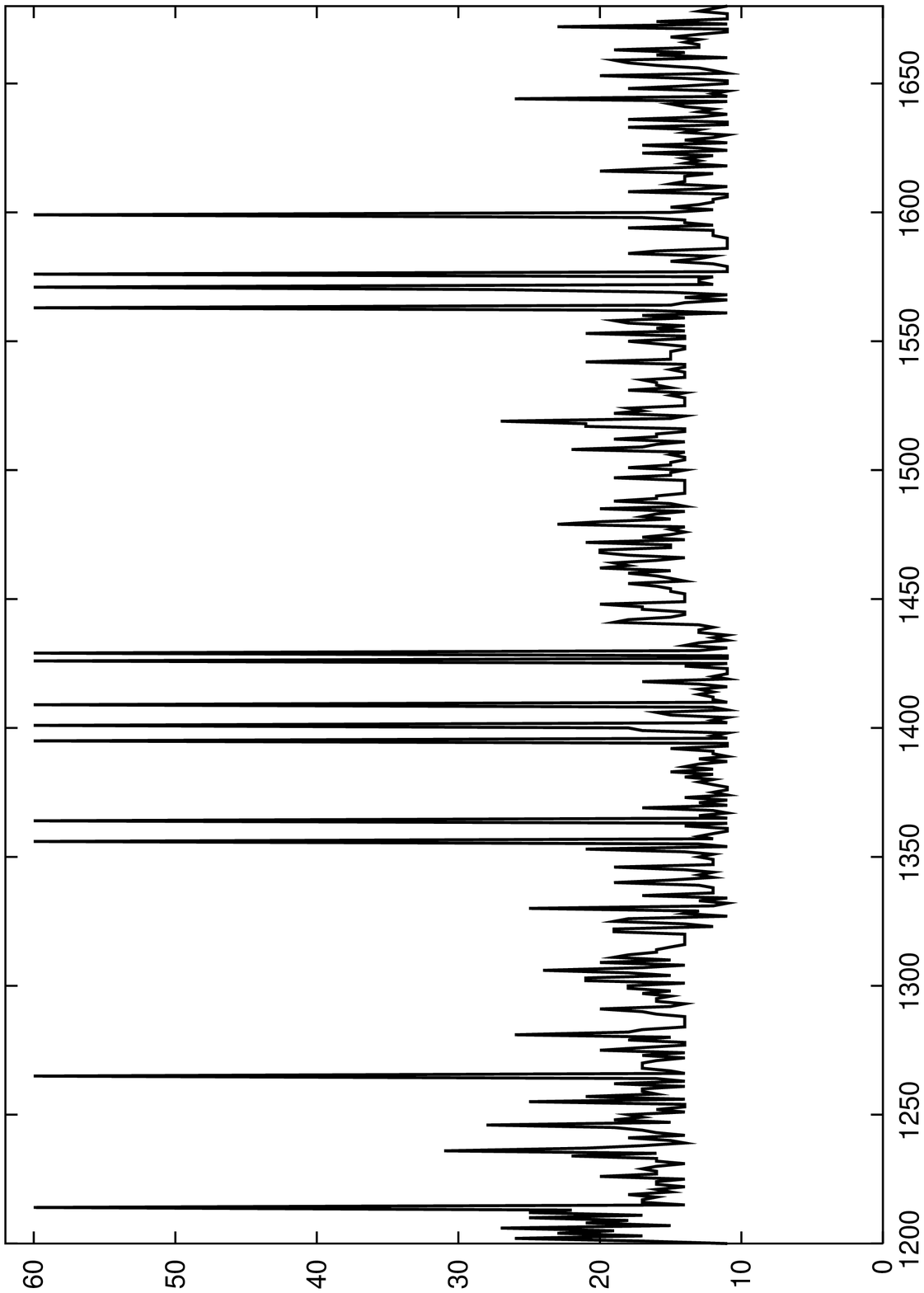}
\includegraphics{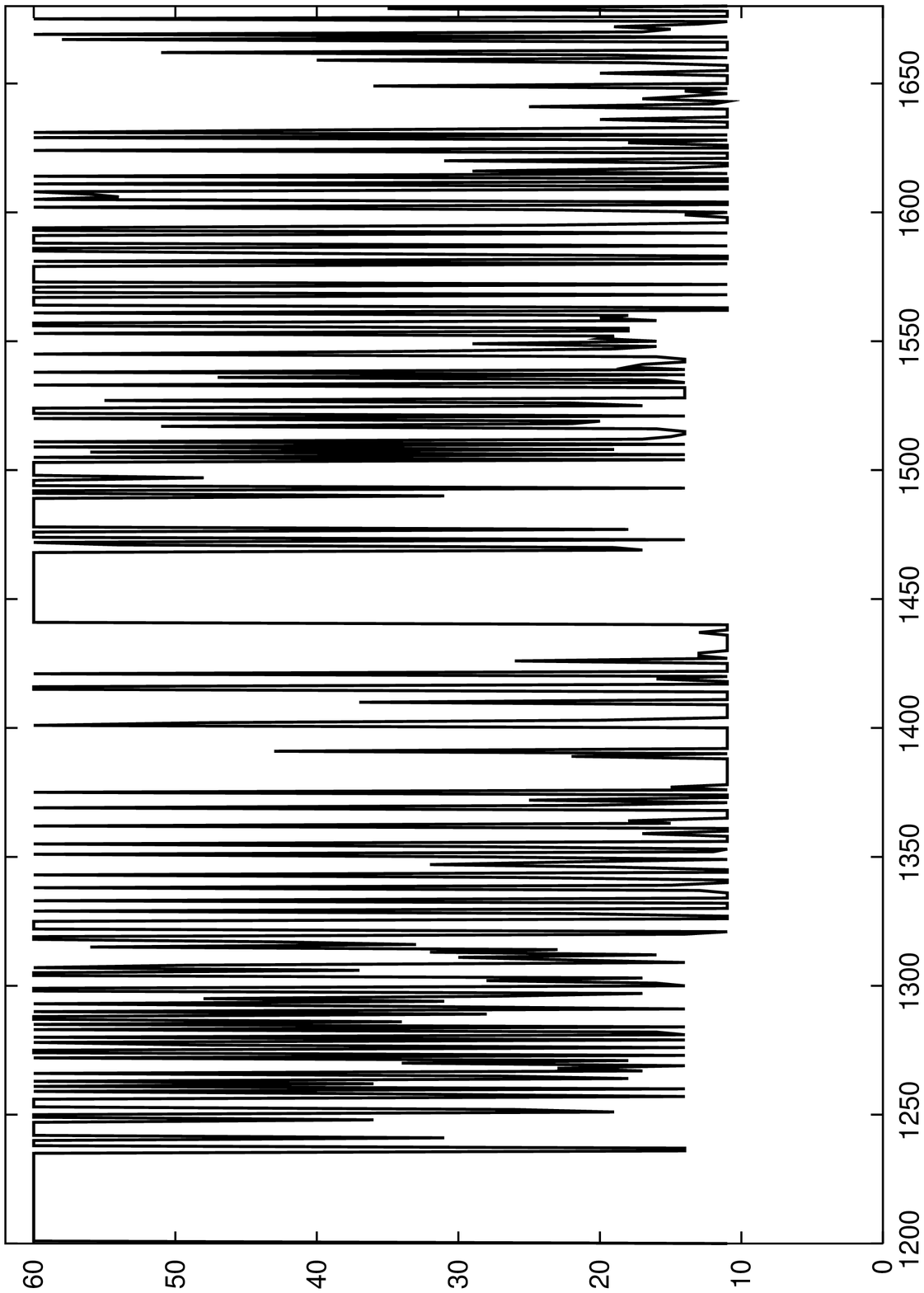}
\caption{Typical behaviour of a simulated robot moving on various
$10\times 11$ boards: the first left plot indicates the grid, the boundary walls, 
the starting position on the bottom line and the goal upper line. 
{\it Left set of plots}: The first three rows
show realizations of paths taken by the robot in three different
environments, ``Empty board" (first row), ``Right hand trap" (second
row), and ``Open trap" (third row) respectively. The different
environments were presented to the robot in three {\em consecutive\/}
trials, i.e., the weights are not reinitialized before a new trial.
Within each row the panels represent from left to right: First run,
early performance, and two different realizations of late performance.
In the simulations a maximal allowed number of steps of 100 is imposed
(after which the run is stopped and maximal penalty assigned). The
fourth row represents two trials on  pairs of configurations related
by a mirror symmetry, ``Traps with clue" (left pair) and ``Traps without
clue" (right pair).  Here 120 runs are performed between switches within
a pair, and the weights are not reinitialized after switches. The cutoff
for the maximal number of steps is imposed at 60 for this experiment.
{\it Right set of plots}: Length of the path taken to the upper line of
the board, plotted run by run for the trials shown in the left set of
plots. First row: ``Empty board" for two settings of the randomness
parameter $\beta$. The length of
the best path is 10 steps in this case. Second row: ``Right hand trap"
(left) and ``Open trap" (right), corresponding to second and third row
respectively in the left set of plots. In both cases the length of the
best path is 13 steps. The dotted (lower) line on the plots indicates
the number of steps lost running against obstacles (this has to be added to
the length of the path to give the total number of steps). Third row:
performance on switching experiments for the symmetric pair of ``Traps
with clue" (left) and the  pair of ``Traps without clue" (right).
The best paths have different lengths for the symmetry-related traps,
viz.  13 and 11 steps  for the left and right member of
a trap-pair, respectively. As can be seen, the performance is much
better if the situation provides a clue (left pair of traps), 
which means that the agent
is able to ``identify" it as such and use it.}

\label{f.art13}
\end{figure}

--- the agent ``identifies" goals, ``realises" the impenetrability 
of obstacles, 
 and ``recognises" clues, such as the small obstacles in the switching
experiment (4-th row in the left part of Fig. \ref{f.art13}) which
 do not themselves obstruct the path, but are correlated with,
 and thus ``announce" the direction in which the larger obstacle
 encountered later on is open.
These  ``top-down" behavioural components are  implemented 
``bottom-up"  using only the simple AR-Hebb rule.

 \par\bigskip
\no {\it Illustration 2}

\begin{figure}[htb]
\vspace{8cm}
\includegraphics{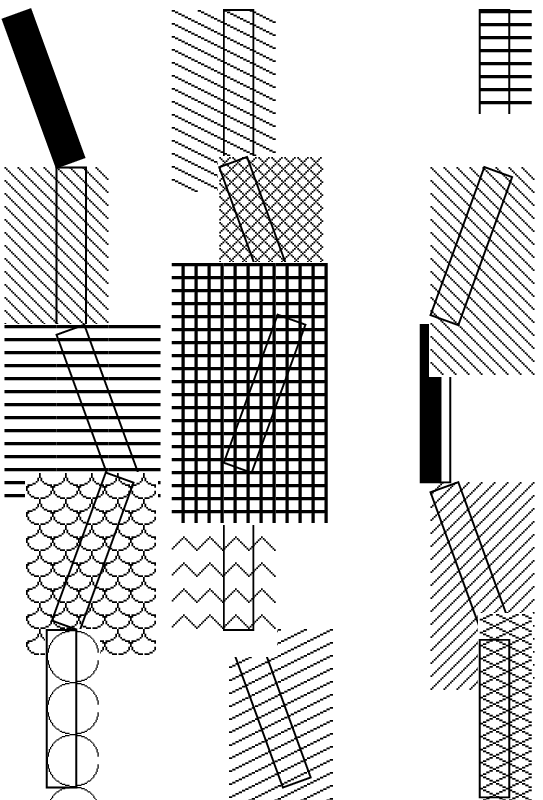}
\includegraphics{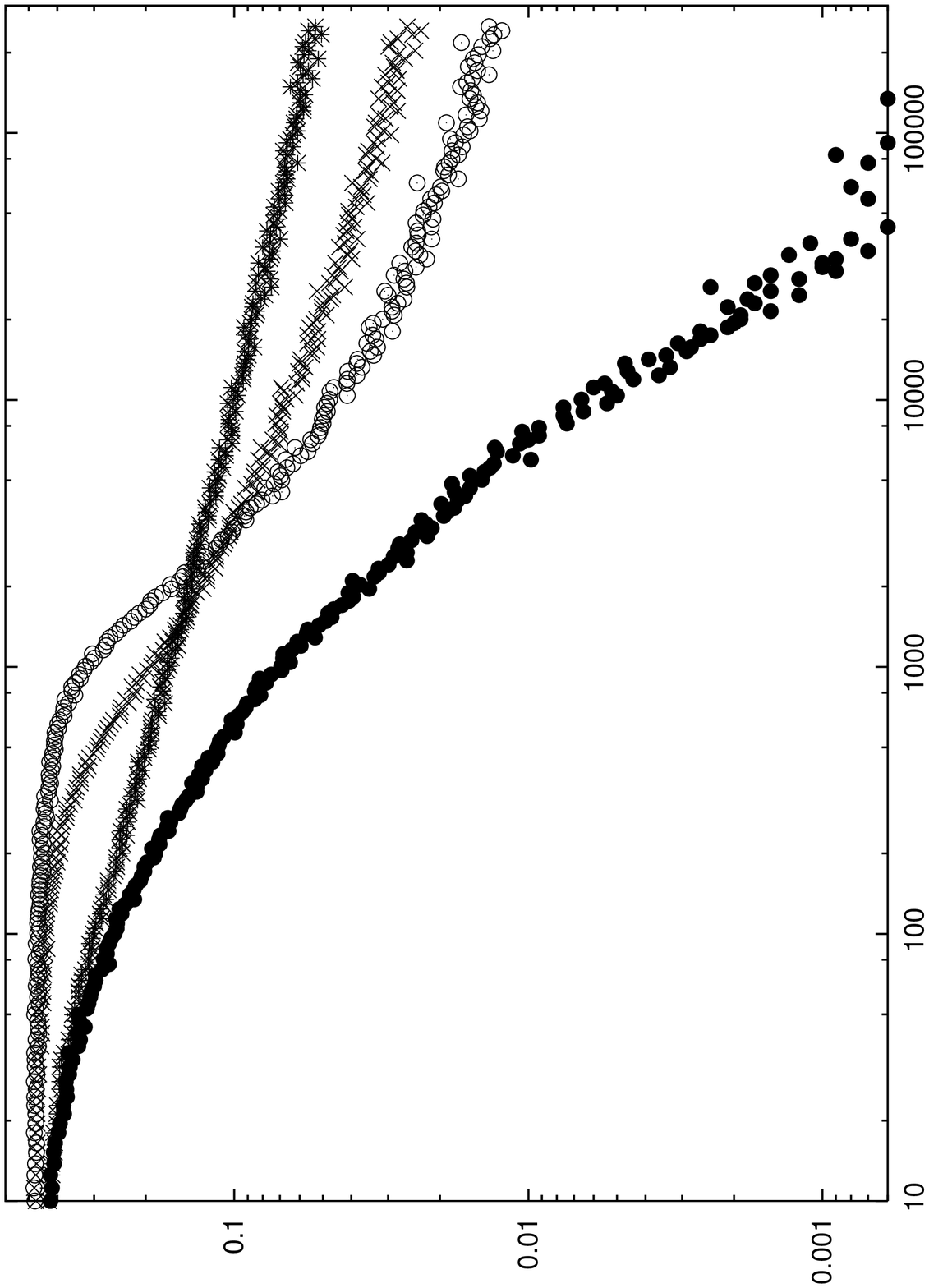}
\caption{Looking for strings of ``black" signature: length of the strings 
$N=20$, length of 
the string sequences $L=5$. $\epsilon_G$ vs $\alpha$ for various starting 
learning steps $1/\sqrt{Q(0)},\ Q(0)= 50000$ (empty circles), 5000 (`x'-s) and 50
(asterisks), and $\lambda$ near the corresponding thresholds 
(0.250, 0.420 and 0.560, respectively).
The black circles represent the decay of the generalization error for the tuning 
$\lambda \propto {\bar e}_q$ ($\lambda$(0)=0.560, $Q(0)=50$).}
\label{f.HI-gen}
\end{figure}

Consider the problem of identifying a certain pattern which is part of a 
sequence of different patterns -- such as a certain gene in a chromosome, 
say. We can only know that the pattern is or is  not 
there -- e.g., from the expected expression of the gene in the phenotype.
A similar problem may be that of finding the scent signature of an (unknown) 
unpalatable 
component in food: The animal tries food in various combination
and can only judge about the combination as a whole.
 In both cases we must allow for variations of the
pattern we are trying to identify (defining thus the class, say, +1). 
For definiteness we take sequences (bags)
of 5 patterns, each pattern being itself a string of $N=20$ bits of information.
 We assume that a ``positive" string  may be contained 
at most once in each sequence (in a bag), and we allow about $10^4$ variations of
it (while there are about $10^6$ patterns of class -1). The student 
is presented with sequences which 
contain a ``positive" string with 50\% probability. The student
only knows (from the observation of the phenotype, from
the effect of the food consumption) whether or not a class +1  
string is contained in the
bag. He or she
therefore makes proposals on the basis of his/her momental knowledge -- 
the synapse strengths --  and then updates the latter globally
taking into account whether its own conclusion matches the observation. 
We therefore have a particular hidden instance problem with structured data 
(HI-SD).  In Fig. \ref{f.HI-gen} we present simulation results for this system
for various initial step sizes $1/\sqrt{Q(0)}$ and parameter
 $\lambda$ chosen to be near the corresponding thresholds. 
 We also  demonstrate that it is possible to improve the algorithm 
by a simple tuning of $\lambda$ with the running average ${\bar e}_q
=\frac{1}{q}\sum_{q'=0}^q e_{q'}$.
The beneficial effect of this tuning can in a certain sense
be seen as reinforcing our claim that it is a {\em combination\/}
of both mechanisms --- the autonomous Hebbian association, and
the synaptic response to evaluative feedback --- which allows
learning under the conditions of non-specific feedback to be
successful. Tuning with the  average 
${\bar e}_q$ is an effective way to ensure that
{\em the balance between both mechanisms\/} is maintained
throughout the learning history, in particular in later stages
with good generalization, where changes via replay become both
small and rare, because $e_q$ will often be small or zero.
 Early and fast learning are thus
easy to achieve: in our illustration already after a few 
seconds of CPU the error drops 
far below 1\%. Again, however, we see the main interest of 
the learning model 
presented here in
its simplicity and versatility in connection with understanding 
biological developments.

\end{document}